\documentclass[zamm,a4paper,fleqn]{w-art}
\usepackage{times}
\usepackage{w-thm}
\usepackage{cite}
\usepackage{amsmath,psfrag}
\usepackage{amssymb}
\usepackage[]{graphicx}
\begin{document}
\DOIsuffix{theDOIsuffix}
\Volume{VV}
\Issue{I}
\Month{MM}
\Year{YYYY}


\Receiveddate{XXXX}
\Reviseddate{XXXX}
\Accepteddate{XXXX}
\Dateposted{XXXX}



\title[Parameter identification in elasto-plasticity with measurement errors]{Parameter identification in
elasto-plasticity: distance between parameters and impact of measurement errors}


\author[A. V. Shutov]{Alexey V. Shutov\inst{1,2}%
  \footnote{Corresponding author,~e-mail:~\textsf{alexey.v.shutov@gmail.com},
            Phone: +07\,383\,333\,17 46,
            Fax: +07\,383\,333\,16 12}}
\address[\inst{1}]{Lavrentyev Institute of Hydrodynamics, pr. Lavrentyva 15, 630090, Novosibirsk, Russia}
\author[A. A.  Kaygorodtseva]{Anastasia A.  Kaygorodtseva\inst{1,2,}}
\address[\inst{2}]{Novosibirsk State University, ul. Pirogova 2, 630090, Novosibirsk, Russia}
\begin{abstract}

A special aspect of parameter identification in finite-strain elasto-plasticity is considered. Namely, we
analyze the impact of the measurement errors on the resulting set of material parameters.
In order to define the sensitivity of parameters with respect to the measurement errors,
a mechanics-based distance between two sets of parameters is introduced. Using this distance function,
we assess the reliability of certain parameter identification procedures. The assessment involves introduction
of artificial noise to the experimental data;
the noise can be both correlated and uncorrelated. An analytical procedure
to speed up Monte Carlo simulations is presented. As a result, a simple tool for estimating
the robustness of parameter identification is obtained. The efficiency of the approach is
illustrated using a model of finite-strain elasto-plasticity, which accounts for
combined isotropic and kinematic hardening. It is shown that dealing with correlated
measurement errors, most stable identification results are obtained for non-diagonal weighting matrix.
At the same time, there is a conflict between the stability and accuracy.

\end{abstract}
\maketitle                   

\section{Introduction }

The classical approach to the identification of material parameters is
based on the minimization of a certain error functional (target function),
which reflects the deviation of simulation results from the available
experimental data \cite{Beck}. This procedure is rather general, but, unfortunately,
does not provide any insight into the ``quality'' of the identified parameters. A practicing engineer might want to know,
how reliable the obtained set of material parameters is.  In order to overcome this problem,
one can analyze the sensitivity of the material parameters with respect to the measurement errors.

A simple tool for assessing the quality of the identification procedure
is correlation matrices:
The identification procedure is considered reliable if the correlations between individual parameters
are separated from $ \pm 1$, see, for example, \cite{Harth2004}, \cite{Harth2007},
\cite{KreissigBenedix}, \cite{ShutovKreissig2010}.
If, in contrast, the correlation between parameters $p_i$ and $p_j$ is close to $ \pm 1$, then a small change in $p_i$  can be
compensated by a suitable change in $p_j$ retaining the same simulation results. In such a situation, parameters
$p_i$ and $p_j$ can not be identified in a reliable way (cf. \cite{Johansson2007}).
The covariance of identified material parameters provides a more detailed information
on the quality of the identification procedure.
In \cite{Harth2004}, a set of experiments is
identified which leads to the smallest sensitivity (in terms of covariance)
of the parameters to the experimental errors.

Obviously, the quality of the parameter identification depends on the
completeness of the experimental data. A big body of information
can be provided by experiments with an inhomogeneous stress-strain state.
Therefore, in a number of publications (cf. \cite{Mahnken1996}, \cite{KreissigBenedix}, \cite{Hartmann2006}, \cite{ShutovLarichkin2017})
the parameter identification is carried out using
finite element method to model experiments with heterogeneous stress-strain distribution.
For two-dimensional problems with measured displacement fields, the virtual field method can be used as well
  with the advantage
that expensive FEM computations are not needed (cf. \cite{Grediac1998}, \cite{Grediac}, \cite{AvrilEtAl2004}, \cite{Rahmani2014}).

In \cite{ShutovKaygorodtseva} the sensitivity of the material parameters with respect to measurement errors
was minimized by an appropriate choice of the weighting coefficients. The main idea behind this procedure is that
some experimental data may be more important than the others.
Another aspect is that the accurately measured experimental data must be granted a
larger weight than the less precise ones.
Thus, in \cite{Beck} the following
recommendation is suggested: If one needs to join two different target functions $\Phi_1$ and $\Phi_2$ pertaining to two different
experiments in a single target function $\Phi$, the following weights should be chosen: $\Phi = \frac{1}{\sigma_1^2} \Phi_1 +
\frac{1}{\sigma_2^2} \Phi_2$, where $\sigma^2_1$ and $\sigma^2_2$ are the variances of the noise in both experiments.
In a more general case, the weighting coefficients can be chosen in such a way as to ensure that the errors in residuals
exhibit equal variance. For instance, having $N$ experimental data $Exp_1$, $Exp_2$, ..., $Exp_N$ with variances
$\sigma^2_1$, $\sigma^2_2$, ..., $\sigma^2_N$ and corresponding
model predictions $Mod_1$, $Mod_2$, ..., $Mod_N$ one may choose the error functional
\begin{equation}\label{Basic}
\Phi = \frac{1}{\sigma_1^2} (Exp_1 - Mod_1)^2 + \frac{1}{\sigma_2^2} (Exp_2 - Mod_2)^2 + ... +
\frac{1}{\sigma_N^2} (Exp_N - Mod_N)^2.
\end{equation}
Note that this formula does not account for the
correlation between measurement errors in different experiments.
A further generalization of this formula will be discussed in Section 5.3.

In order to decide on which identification strategy is most insensitive
to the measurement errors, the sensitivity must be measured in numbers.
A straightforward approach, based on the sensitivity of \emph{individual} material parameters $p_i$, $i=1,...,n$,
does not allow to grasp the \emph{collective} behaviour of the parameter set $\vec{p} = (p_1,...,p_n)^{\text{T}}$.
Thus, a metric in the space of material parameters is needed which allows
one to measure a distance between two sets of material parameters.
In the current paper, a mechanics-based metric in the space of material parameters is proposed.
This metric is advantageous over the conventionally used Euclidean metric ($l_2$-metric).
In particular, the mechanics-based metric in invariant under re-parametrization of the material model;
situation where different material parameters are of different dimension does not pose any problem for this metric.
Loosely speaking, when working with the mechanics-based metric the impact of each parameter is
proportional to its influence on the stress response.

The paper is organized as follows. In Section 2, a general procedure for finding material
parameters is discussed, which is based on the minimization of a certain least-square error functional.
In Section 3, we present a short overview of different stochastic models of noise used in reliability analysis and
a simple solution for linearized model response is presented. In Section 4,
the announced physics-based metric is defined. A series of demonstration
problems of parameter identification is solved in Section 5. Finally, in Section 6
the main results are summarized and discussed.

\section{A general procedure for finding material parameters}

Assume that in a certain experimental program a set of $N$ experimental observations is available. Here, the result of each observation
is a certain real number.
Denote by $\overrightarrow{\text{Exp}} = (Exp_1, Exp_2, ..., Exp_N)^{\text{T}}$ the corresponding vector of experimental data.
For a given physical model, the corresponding theoretical predictions
are denoted by $\overrightarrow{\text{Mod}} = (Mod_1, Mod_2, ..., Mod_N)^{\text{T}}$. In a standard setting,
these theoretical values depend on $n$ real-valued parameters $p_1$, $p_2$, ..., $p_n$.
We write for brevity $\vec{p} = (p_1, p_2,...,p_n)^{\text{T}}$.
Obviously, the number of parameters should be smaller than the number of experimental results: $n < N$.

\textbf{Remark 1.} In some applications it is reasonable
to impose restrictions on the set of material parameters.
Some of the restrictions represent algebraic equations
 of type $g_1(\vec{p}) = 0$, ... ,
 $g_k(\vec{p}) = 0$. Restrictions of another type
 are given by inequalities $h_1(\vec{p}) > 0$, ...,
 $h_l(\vec{p}) > 0$. Here, $k$ and $l$
 is the number of equality constraints and inequality constraints, respectively.
Equality constraints are not consider in the current study; in some cases
they can even be used to regularize the identification
procedure by reducing the number of material parameters \cite{ShutovKreissig2010}.
 As for inequality constraints, some authors suggest that
 a ``good'' identification procedure must satisfy
 these constraints in a natural way
(cf. the discussion in \cite{KreissigBenedix}).
Therefore, they are not introduced in the current setting as well.
   $\blacksquare$

Let $\mathbf{W}$ be a given square $N \times N$ matrix; assume that it is symmetric and positive
definite. Usually, the parameter identification is reduced to the following optimization (minimization) problem
\begin{equation}\label{OptimizProblem}
\vec{p} = \text{argmin} (\Phi(\vec{p})), \quad
\Phi(\vec{p}) = \overrightarrow{\text{Resid}}^{\text{T}} \ \mathbf{W} \ \overrightarrow{\text{Resid}}, \quad
\overrightarrow{\text{Resid}} = \overrightarrow{\text{Exp}} - \overrightarrow{\text{Mod}}.
\end{equation}
Here, $\overrightarrow{\text{Resid}}$ is the so-called residuum, being seen as a deviation
of theoretical results from the experimental data.
This optimization problem is equivalent to the minimization
of the $l_2$-norm of a modified residuum $\overrightarrow{\text{WResid}}$:
\begin{equation}\label{OptimizProblem2}
\vec{p} = \text{argmin} (\Phi(\vec{p})), \quad
\Phi(\vec{p}) = \| \overrightarrow{\text{WResid}} \|^2 = \overrightarrow{\text{WResid}}^{\text{T}} \ \overrightarrow{\text{WResid}}, \quad
\overrightarrow{\text{WResid}} := \ \mathbf{W}^{1/2} \ \overrightarrow{\text{Resid}}.
\end{equation}
In contrast to \eqref{OptimizProblem}, problem \eqref{OptimizProblem2} can be solved using standard procedures, like the well-established
and reliable
Levenberg-Marquardt method \cite{LevenbergMarquardt}.
Some considerations regarding a ``good'' matrix $\mathbf{W}$ will be presented in Section 5.3.

\section{Introduction of noise to experimental stress-strain curves }

\subsection{Types of noise}

It is natural to assume that the available experimental data $\overrightarrow{\text{Exp}} = (Exp_1, Exp_2, ..., Exp_N)^{\text{T}}$
 are contaminated by measurement errors. In other words,
in reality, noisy data are available.
Usually one assumes that the error is additive \cite{Beck}:
\begin{equation}\label{ErrorIsAdditive}
Noisy \ data_i = Exp_i + Noise_i.
\end{equation}

\begin{figure}\centering
\scalebox{0.80}{\includegraphics{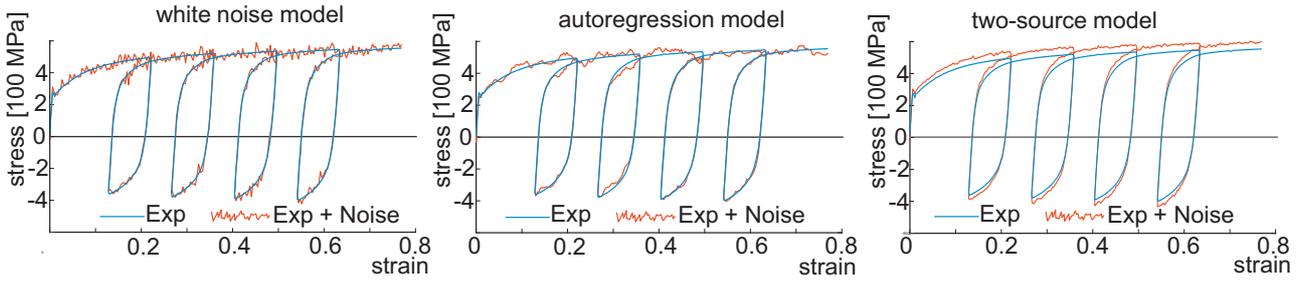}}
\caption{Three types of noise: white noise (left), autoregression model (middle),
two-source model (right). \label{fig1}}
\end{figure}

In order to analyze the dependence of the parameter identification procedure on this measurement error, we need a stochastic model of
$Noise_i$. The most simple model of noise is given by the assumption that the measurement errors are \emph{independent} random variables
with a normal distribution with a zero mean and standard deviation $\sigma$ (white noise, see Fig. \ref{fig1} (left))
\begin{equation}\label{WhiteNoise}
Noisy \ data_i \in \mathcal{N} (\mu,\sigma^2), \quad \mu = 0 .
\end{equation}

The white-noise-model is a good choice in many practical situations due to the central limit theorem of the probability theory. The central limit theorem
states that, if a large number of independent random variables is added, their normalized sum converges
to a random variable with a normal distribution. Therefore, the white noise assumption
is used in a number of studies to assess the stability of a
certain identification procedure (cf. \cite{Grediac}, \cite{AvrilEtAl2004}, \cite{ShutovKaygorodtseva})

Another stochastic description of noise is provided by the so-called autoregressive model (AR-model).
Using the AR-model, it is possible to account for impact of previous measurement errors
on the next measurement.
For instance, assuming the Yule-Walker equations, the following
scheme is obtained (see, for instance, \cite{Seibert}, \cite{Harth2004}, \cite{Harth2007})
\begin{equation}\label{ARmodel}
Noise_i = \alpha Noise_{i-1} + \varepsilon_i, \quad \alpha \in [0,1),
\end{equation}
where $\varepsilon_i$ is a sequence of independent normally distributed random variables: $\varepsilon_i \in \mathcal{N} (0,\sigma^2)$;
$\alpha$ is the autoregression parameter.
According to \eqref{ARmodel}, the errors are not independent random variables, but correlated in a certain way (see Fig. \ref{fig1} (middle)).
Such a correlation can arise due to inertia effects in the testing equipment. For
example, in experiments involving the occurrence of the L\"uders bands, one may assume
a noise caused by resonance in the load cell \cite{Avril2008}.

Another stochastic model of noise
which we call ``two-source model'' is as follows.
We assume that two independent sources of noise are active, responsible respectively for correlated and non-correlated errors.
Thus we have
\begin{equation}\label{TwoSourseModel}
Noise_i = Noise^{non-correlated}_i + Noise^{correlated}_i.
\end{equation}
Here $Noise^{non-correlated}_i$ corresponds to the previously mentioned white-noise. Thus,
it is a sequence of independent normally distributed variables with the standard deviation $\sigma_1$:
$Noise^{non-correlated}_i \in \mathcal{N} (0,\sigma_1^2)$.
For the correlated part of the noise we set
\begin{equation}\label{CorrelNoise}
Noise^{correlated}_i = \varepsilon \frac{Exp_i}{\max\limits_{j} | Exp_j |}, \quad \varepsilon \in \mathcal{N} (0,\sigma_2^2).
\end{equation}
Dealing with experimental identification of stresses, the correlated error \eqref{CorrelNoise} can be caused by a
wrong calibration of dynamometer or a wrong measurement of the sample cross-section (see Fig. \ref{fig1} (right)).
The ``two-source model''
will be employed for Monte Carlo computations in Section 5.3.

In some studies, the sensitivity of the material parameters is analyzed
 assuming that the noise is uniformly distributed in a certain interval \cite{Johansson2007}.
This stochastic model is implemented not due to ist plausibility, but rather due
to its simplicity.
Another stochastic model may consider additional
noise caused by a play in the testing assembly \cite{ShutovKaygorodtseva}.
Note that this section refers solely to the stress-strain curves.
Dealing with experimentally measured fields like the displacements obtained
by digital image correlation,
more sophisticated techniques are needed (cf. \cite{Rosic2013}).
In the more general context of reliability analysis, one may consider the Weibull or lognormal distribution
of noise. Typically, this is done for positive strength parameters like elasticity modulus
or yield stress of the material (cf. \cite{Kaminski}).

\subsection{Fast optimization for linearized response}

Let us consider the optimization problem \eqref{OptimizProblem2} with the vector of experimental data $\overrightarrow{\text{Exp}}$.
For a given matrix $\mathbf{W}$, the minimizing set of parameters $\vec{p}_*$ and the Jacobian of the model response are as follows
\begin{equation}
\vec{p}_* = \arg \min \Phi, \quad \mathbf{J} = \frac{\partial \overrightarrow{\text{Mod}(\vec{p})}} {\partial \vec{p}}|_{\vec{p}_*}.
\end{equation}
Along with this basic (unperturbed) optimization problem we consider a case where the measurements are contaminated by a small noise.
In order to build an analytical procedure, the model response function $\overrightarrow{\text{Mod}(\vec{p})}$ is linearized near $\vec{p_*}$:
\begin{equation}\label{Linearization}
\overrightarrow{\text{Mod}(\vec{p})} \approx  \overrightarrow{\text{Mod}^{lin}(\vec{p})} = \overrightarrow{\text{Mod}(\vec{p}_*)} + \mathbf{J}(\vec{p} - \vec{p}_*).
\end{equation}
The error functional, which corresponds to the optimization problem with noisy data, takes the form
\begin{equation}\label{ErrorFuncLin}
\Phi^{noisy}(\vec{p}) = \overrightarrow{\text{Resid}}^{\text{T}} \ \mathbf{W} \ \overrightarrow{\text{Resid}}, \quad
\overrightarrow{\text{Resid}} = \overrightarrow{\text{Exp}} + \overrightarrow{\text{Noise}} - \overrightarrow{\text{Mod}^{lin}} =
\overrightarrow{\text{Exp}} + \overrightarrow{\text{Noise}} -  \overrightarrow{\text{Mod}(\vec{p}_*)} - \mathbf{J}(\vec{p} - \vec{p}_*).
\end{equation}
This error functional can be re-written in the following form
\begin{equation}\label{ErrorFuncLin2}
\Phi^{noisy}(\vec{p}) = \overrightarrow{\text{WResid}}^{\text{T}} \ \overrightarrow{\text{WResid}}, \quad
\overrightarrow{\text{WResid}} := \ \mathbf{W}^{1/2} \ \overrightarrow{\text{Resid}}.
\end{equation}
We introduce the following abbreviation
\begin{equation}\label{NotationA}
\overrightarrow{A} := \mathbf{W}^{1/2} \ \big( \overrightarrow{\text{Exp}} + \overrightarrow{\text{Noise}} - \ \overrightarrow{\text{Mod}(\vec{p}_*)} + \mathbf{J} \ \vec{p}_*   \big).
\end{equation}
Thus, the error functional is a quadratic function of $\vec{p}$
\begin{equation}\label{ErrorFuncLin3}
\Phi^{noisy}(\vec{p}) = \big(\overrightarrow{A} - \mathbf{W}^{1/2} \ \mathbf{J} \vec{p}\big)^{\text{T}} \big(\overrightarrow{A} - \mathbf{W}^{1/2} \ \mathbf{J} \vec{p}\big).
\end{equation}
Its derivative with respect to $\vec{p}$ is given by the linear function
\begin{equation}\label{DerivativeOfPhi}
\frac{\displaystyle \partial \Phi^{noisy}(\vec{p})}{\displaystyle \partial \vec{p} } = - 2
 \big( \overrightarrow{A}  - \mathbf{W}^{1/2} \ \mathbf{J} \vec{p} \big)^{\text{T}}  \  \mathbf{W}^{1/2} \ \mathbf{J}.
\end{equation}
Since $\vec{p}$ is a local extremum of $\Phi^{noisy}(\vec{p})$, this derivative is equal to zero.
After some rearrangements we arrive at the analytical solution
\begin{equation}\label{AnalytSolution}
\vec{p} = \big( \mathbf{J}^{\text{T}} \mathbf{W} \mathbf{J}  \big)^{-1} \big(\mathbf{W}^{1/2} \ \mathbf{J} \big)^{\text{T}} \ \overrightarrow{A}.
\end{equation}
Note that in the noise-free case (when $\overrightarrow{\text{Noise}} = \overrightarrow{0} $), the
original solution is restored, thus yielding $\vec{p} = \vec{p}_*$.

\section{How to measure a distance between two sets of parameters}

Having two sets of material parameters, $\vec{p}^{\ (1)}$ and $\vec{p}^{\ (2)}$, one needs
to estimate the distance between them. Such a distance $\text{dist}(\vec{p}^{\ (1)},\vec{p}^{\ (2)})$
should be small if the parameter sets are close to each other in a certain sense. In the following sections, this function
will be used to estimate the dependence of material parameters on the experimental errors.

\subsection{Euclidean norm}

One of the simplest ways to measure a distance is to
employ the Euclidean norm
\begin{equation}\label{EuclideanDistance}
\text{dist}^{\text{Euclidean}}(\vec{p}^{\ (1)},\vec{p}^{\ (2)}) := \sqrt{\sum_{i=1}^{n} \big( p^{(1)}_i - p^{(2)}_i \big)^2}.
\end{equation}
Unfortunately, in most practical situations, this norm \emph{is physically absurd} since
the parameter set $\vec{p}$ encapsulates quantities with different physical dimensions.
In order to resolve this issue, a non-dimensional Euclidean norm can be used (see, for instance, \cite{ShutovKaygorodtseva})
\begin{equation}\label{EuclideanDistanceNondim}
\text{dist}^{\text{Eucl. non-dimens.}}(\vec{p}^{\ (1)},\vec{p}^{\ (2)}) := \sqrt{\sum_{i=1}^{n} \Big(\frac{ p^{(1)}_i - p^{(2)}_i}{p^{*}_i} \Big)^2},
\end{equation}
where $p^{*}_i$ is a typical (characteristic) value of the parameter $p_i$.
Obviously, there is a certain arbitrariness in the choice of $p^{*}_i$ and substantial
difficulties will arise when $p^{*}_i = 0$.

For a number of complex material models certain material parameters \emph{can not be identified}
with a suitable accuracy. At the same time, large variance in these bad trackable parameters
does not have any substantial impact on the overall stress response \cite{Zhang1995}.
Thus, an essential drawback of the Euclidean norm is that all the parameters
$\frac{\displaystyle p_i}{\displaystyle p^{*}_i}$
are treated by \eqref{EuclideanDistanceNondim} in the same way, regardless of their
importance for the physical problem under consideration.

Moreover, the Euclidean metric is sensitive to a re-paramterization of the material model.
More precisely, when the parameter set $\vec{p}$ is a one-to-one function of some new parameters $\vec{\rho}$
( $\vec{p} =  \vec{p} (\vec{\rho})$ ), then $\text{dist}\big(\vec{p}(\vec{\rho}^{\ (1)}),\vec{p}(\vec{\rho}^{\ (2)})\big) \neq
\text{dist}(\vec{\rho}^{\ (1)},\vec{\rho}^{\ (2)})$.
In order to resolve these problems, mechanical considerations are needed.

\subsection{Mechanics-based metric in the space of material paramters}

Dealing with finite-strain elasto-plasticity,
it is reasonable to introduce a strain-controlled loading path at a certain material point.
Let $T$ be the time of the loading process, $\mathbf{F}(t)$ be the
deformation gradient tensor given as a function of time $t \in [0,T]$.
Assuming  a simple material of Noll's type \cite{Noll1972},
the local history of the Cauchy stress tensor $\mathbf{T}(t)$ is a unique function
of the local deformation history $\mathbf{F}(t)$ and material parameters
$\vec{p} = (p_1, p_2,...,p_n)^{\text{T}}$:\footnote{For so-called ``simple materials with
initial conditions'' \cite{Noll1972}, the stress history may depend on the initial values of the
internal variables. The initial values can be included in the set of material parameter $\vec{p}$.}
\begin{equation}\label{MaterialModelItself}
\mathbf{T}(t,\vec{p}) = \mathop{ \mathbf{T} }_{0 \leq t' \leq t}  \big(  \mathbf{F}( t'), \vec{p} \big), \quad
\text{for all} \ t \in [0,T].
\end{equation}
In the left-hand side of this relation, the dependence of the stress response on
the deformation history $\mathbf{F}(t)$ is omitted for brevity.
The deformation history $\mathbf{F}(t)$ and the material model \eqref{MaterialModelItself}
uniquely define a mechanics-based distance between two sets of material
parameters  $\vec{p}^{\ (1)}$ and $\vec{p}^{\ (2)}$ as
\begin{equation}\label{PhysBasedDist}
\text{dist}^{\mathbf{F}}(\vec{p}^{\ (1)}, \vec{p}^{\ (2)}) := \max\limits_{t \in [0,T]}
\| \mathbf{T}(t,\vec{p}^{\ (1)}) - \mathbf{T}(t,\vec{p}^{\ (2)})  \|.
\end{equation}
In order to confirm that $\text{dist}^{\mathbf{F}}(\vec{p}^{\ (1)}, \vec{p}^{\ (2)})$ defines a metric
on a certain set of material parameters, one needs to check the following conditions:
\begin{equation}\label{Cond1}
(i): \quad \text{dist}^{\mathbf{F}}(\vec{p}^{\ (1)}, \vec{p}^{\ (2)}) \geq 0,
\end{equation}
\begin{equation}\label{Cond2}
(ii): \quad \text{dist}^{\mathbf{F}}(\vec{p}^{\ (1)}, \vec{p}^{\ (2)}) = 0
\quad \text{if and only if} \quad \vec{p}^{\ (1)} = \vec{p}^{\ (2)},
\end{equation}
\begin{equation}\label{Cond3}
(iii): \quad \text{dist}^{\mathbf{F}}(\vec{p}^{\ (1)}, \vec{p}^{\ (2)})
= \text{dist}^{\mathbf{F}}(\vec{p}^{\ (2)}, \vec{p}^{\ (1)}),
\end{equation}
\begin{equation}\label{Cond4}
(iv): \quad \text{dist}^{\mathbf{F}}(\vec{p}^{\ (1)}, \vec{p}^{\ (3)}) \leq
\text{dist}^{\mathbf{F}}(\vec{p}^{\ (1)}, \vec{p}^{\ (2)}) +
\text{dist}^{\mathbf{F}}(\vec{p}^{\ (2)}, \vec{p}^{\ (3)}).
\end{equation}
Conditions $(i)$, $(iii)$, and $(iv)$ are trivially satisfied.
Condition $(ii)$ is satisfied only if the loading programm $\mathbf{F}(t)$
makes each material parameter ``visible''.
In the purely elastic range, two different sets of hardening parameters
may produce the same stress response, thus yielding a zero distance between
these sets of parameters. In order to avoid this undesired effect, the prescribed strains
must be large enough. Concrete examples will be considered in Section 5.3.

\textbf{Remark 2.} Definition \eqref{PhysBasedDist} is based on a local strain history.
Obviously, some other practice-related distances can be defined
using a solution of a practical boundary value problem.
On the other hand, as will be shown in Section 5.3, a concrete choice of the
local loading history is not so important and different loading histories yield similar results.
$\blacksquare$

\section{Illustration problem: model with combined isotropic-kinematic hardening}

\subsection{Experimental data for the steel 42CrMo4}

For demonstration purposes we consider the parameter identification problem, basing
on the experimental data, reported in \cite{ShutovKuprin} for the steel 42CrMo4.
During testing, thin-walled tubular specimens were subjected to non-monotonic torsion.
The measured shear stresses are plotted versus the shear strain in Fig. \ref{fig2} (top left).
As can be seen from the figure, the material exhibits a strong Bauschinger effect coupled to expansion of the elastic domain.
In order to describe this type of stress response, finite-strain plasticity models with a combined isotropic-kinematic hardening
are usually implemented. The presented measurement results will serve as a basis
for the identification of material parameters.
As discussed in \cite{ShutovKuprin}, the initial (as-received) state can be idealized as isotropic.
This observation is important for the identification of the initial state.

\subsection{Deterministic plasticity model of Shutov and Krei\ss ig (2008)}

The model of finite-strain viscoplasticity proposed by Shutov and  Krei\ss ig  (cf. \cite {ShutovKreissig2008})
is formulated in a geometrically exact manner. The description of the nonlinear kinematics
is based on the nested split of the deformation gradient tensor, originally proposed by Lion in \cite{Lion}.
Relations between stresses and elastic strains are of hyperelastic type (cf. \cite{Haupt}).
This combination of constitutive assumptions is shown to have numerous advantages over competing alternatives \cite{ShutovAnalysisOfSome}.
The model accounts for nonlinear isotropic and kinematic hardening, it is objective and thermodynamically consistent, it is free
from spurious shear stress oscillations under simple shear, and it is (weakly) invariant under
isochoric changes of the reference configuration \cite{ShutovPfeiffer}, \cite{ShutovAnalysisOfSome}. Some micromechanical arguments in favour of the nested
multiplicative split are presented in \cite{Shutov2018MSCF}.

The deformation at a material point is captured by the right Cauchy-Green tensor $\textbf{C} := \mathbf{F}^{\text{T}}\mathbf{F}.$
 The current state of the material is described by internal variables of the right Cauchy-Green type:
$\textbf{C}_{\text{i}}$ for inelastic strains and
$\textbf{C}_{\text{i1}}$, $\textbf{C}_{\text{i2}}$ for the inelastic strains of substructure.
Tensors $\textbf{C}$, $\textbf{C}_{\text{i}}$, $\textbf{C}_{\text{i1}}$, and $\textbf{C}_{\text{i2}}$ are symmetric and positive definite.
Additionally, two scalar-valued internal variables are employed: accumulated inelastic arc-length (Odqvist parameter) $s$ and its
dissipative part $s_{\text{d}}$.

By $\psi$ we denote the Helmholz free energy per unit mass. Assume that it is decomposed into the following summands (cf. \cite{ShutovKuprin}):
\begin{equation}
\psi = \psi_{\text{el}}(\textbf{C}\textbf{C}_{\text{i}}^{-1}) + \psi_{\text{kin}1}(\textbf{C}_{\text{i}}\textbf{C}_{\text{i1}}^{-1}) + \psi_{\text{kin}2}(\textbf{C}_{\text{i}}\textbf{C}_{\text{i2}}^{-1}) + \psi_{\text{iso}}(s-s_{\text{d}}).
\end{equation}
Here, $\psi_{\text{el}}(\textbf{C}\textbf{C}_{\text{i}}^{-1})$ captures the energy storage due to
macroscopic elastic deformations;  $\psi_{\text{kin}1}(\textbf{C}_{\text{i}}\textbf{C}_{\text{i1}}^{-1})$,
 $\psi_{\text{kin}2}(\textbf{C}_{\text{i}}\textbf{C}_{\text{i2}}^{-1})$ and $\psi_{\text{iso}}(s-s_{\text{d}})$ are parts of the energy
stored in defects of crystal lattice, they are related to
kinematic and isotropic hardening.
Important limitation of the approach is that the functions
$\psi_{\text{el}}$, $\psi_{\text{kin}1}$, and $\psi_{\text{kin}2}$ are isotropic.
For practical computations we use the following constitutive assumptions:
\begin{equation}
\rho_{\text{R}}\psi_{\text{el}}(\textbf{A}) = \frac{k}{2}(\ln\sqrt{\det\textbf{A}})^2 +\frac{\mu}{2}(\text{tr}\overline{\textbf{A}} - 3),
\end{equation}
\begin{equation}
\rho_{\text{R}}\psi_{\text{kin}1}(\textbf{A}) = \frac{c_1}{4}(\text{tr}\overline{\textbf{A}} - 3), \quad \rho_{\text{R}}\psi_{\text{kin}2}(\textbf{A}) = \frac{c_2}{4}(\text{tr}\overline{\textbf{A}} - 3),
\end{equation}
\begin{equation}
\rho_{\text{R}}\psi_{\text{iso}}(s_{\text{e}}) = \frac{\gamma}{2}(s_{\text{e}})^2,  \quad \overline{\textbf{A}} := (\det\textbf{A})^{-1/3}\textbf{A},
\end{equation}
for any second-rank tensor $\textbf{A}$ and scalar $s_{\text{e}}$. Here,
$k$, $\mu$, $c_1$, $c_2$, $\gamma$ are material parameters; $\rho_{\text{R}}$ denotes the mass
density in the reference configuration.
Employing the standard Coleman-Noll procedure, the second Piola-Kirchhoff stress $\widetilde{\mathbf{T}}$ is related to strains through
\begin{equation}
\widetilde{\mathbf{T}} = 2\rho_{\text{R}} \frac{\partial\psi_{\text{el}}(\textbf{C}\textbf{C}_{\text{i}}^{-1})}{\partial \textbf{C}}|_{\textbf{C}_{\text{i}} = const}.
\label{2ndPK}
\end{equation}
Two backstresses $\widetilde{\mathbf{X}}_1$ and $\widetilde{\mathbf{X}}_2$ and the overall backstress $\widetilde{\mathbf{X}}$
are used in the current paper to capture the translation
of the yield surface in the stress space. These tensors
operate on the reference configuration; they are computed through
\begin{equation}
\widetilde{\mathbf{X}}_1 = 2\rho_{\text{R}} \frac{\partial\psi_{\text{kin}1}(\textbf{C}_{\text{i}}\textbf{C}_{1\text{i}}^{-1})}{\partial \textbf{C}_{\text{i}}}|_{\textbf{C}_{1\text{i}} = const}, \quad
\widetilde{\mathbf{X}}_2 = 2\rho_{\text{R}} \frac{\partial\psi_{\text{kin}2}(\textbf{C}_{\text{i}}\textbf{C}_{2\text{i}}^{-1})}{\partial \textbf{C}_{\text{i}}}|_{\textbf{C}_{2\text{i}} = const}, \quad \widetilde{\mathbf{X}} = \widetilde{\mathbf{X}}_1 + \widetilde{\mathbf{X}}_2.
\end{equation}
A hardening variable $R \in \mathbb{R}$, which is responsible for isotropic expansion of the yield surface, is related to scalar-valued internal
variables:
\begin{equation}
R = \rho_{\text{R}} \frac{\partial\psi_{\text{iso}}(s-s_{\text{d}})}{\partial s}|_{s_{\text{d}} = const}.
\end{equation}
For viscoplastic models, stress states beyond the elastic domain are possible.
The corresponding viscous overstress $f$ depends on the applied strain rate; it is defined by
\begin{equation}
f := \mathfrak{F} - \sqrt{\frac{2}{3}}(K+R), \quad
\mathfrak{F} := \sqrt{ \text{tr}[(\textbf{C} \widetilde{\textbf{T} } -\textbf{C}_{\text{i}}\widetilde{\textbf{X}})^{\text{D}}]^2 },
\end{equation}
where $K$ is the initial quasi-static yield stress,
$(\cdot)^{\text{D}}$ stands for the deviatoric part of a tensor,
$\mathfrak{F}$ is the driving force of the viscoplastic flow.
An inelastic multiplier  $\lambda_{\text{i}}$ is introduced which equals the norm of the inelastic strain rate;
$\lambda_{\text{i}}$ is computed employing the Perzyna law of viscoplasticity
\begin{equation}
\lambda_{\text{i}} = \frac{1}{\eta}\Big\langle\frac{f}{k_0}\Big\rangle^m, \quad \langle x \rangle := \max(x,0).
\label{Perz}
\end{equation}
Here, $\eta$ and $m$ are respectively the viscosity and the stress exponent; $k_0$ is set equal to 1 MPa in order
to obtain a non-dimensional quantity in the bracket.
The evolution of the internal variables is specified by the following constitutive equations
\begin{equation}\label{f3}
\dot{\textbf{C}_{\text{i}}} = 2\frac{\lambda_{\text{i}}}{\mathfrak{F}}(\textbf{C}\widetilde{\textbf{T}} -\textbf{C}_{\text{i}}\widetilde{\textbf{X}})^{\text{D}}\textbf{C}_{\text{i}},
\end{equation}
\begin{equation}\label{f4}
\dot{\textbf{C}}_{1\text{i}} = 2\lambda_{\text{i}}\varkappa_1(\textbf{C}_{\text{i}}\widetilde{\textbf{X}}_1)^{\text{D}}\textbf{C}_{1\text{i}}, \quad
\dot{\textbf{C}}_{2\text{i}} = 2\lambda_{\text{i}}\varkappa_2(\textbf{C}_{\text{i}}\widetilde{\textbf{X}}_2)^{\text{D}}\textbf{C}_{2\text{i}},
\end{equation}
\begin{equation}
\dot{s} = \sqrt{\frac{2}{3}}\lambda_{\text{i}}, \quad \dot{s}_{\text{d}} =\frac{\beta}{\gamma}\dot{s}R.
\end{equation}
Here, $\varkappa_1$, $\varkappa_2$ are parameters governing the saturation of the
kinematic hardening; $\beta$ is responsible for the saturation of the isotropic hardening;
$\dot{(\cdot)}$ is the material time derivative (differentiation with respect to the time $t$
while the particle is held fixed).
In the current study we assume that
the initial state is isotropic, undeformed, and stress free. This yields the following initial conditions
\begin{equation}
\textbf{C}_{\text{i}}|_{t=0} = \textbf{C}_{1\text{i}}|_{t=0} = \textbf{C}_{2\text{i}}|_{t=0} = \textbf{1},
\quad  s|_{t=0} = s_{\text{d}}|_{t=0} = 0.
\end{equation}

Differential equations (\ref{f3}) and (\ref{f4}) describe an incompressible flow:
\begin{equation}
\det(\textbf{C}_{\text{i}}) = \det(\textbf{C}_{1\text{i}}) = \det( \textbf{C}_{2\text{i}}) = 1.
\end{equation}

Robust and efficient
numerical procedures for the case where $\psi_{\text{kin}1}$ and
$\psi_{\text{kin}2}$ are of neo-Hookean type are presented in \cite{Shutov2016}. The case where
$\psi_{\text{kin}1}$ and $\psi_{\text{kin}2}$ are of Mooney-Rivlin type can be dealt with using
an explicit update formula from \cite{Shutov2017}.
The model is implemented into the nonlinear FEM-code MSC.MARC.
Practical applications were analyzed using this model in \cite{Scherzer1}, \cite{Scherzer2}.
Note that the material model summarized in this section is deterministic.
The reader interested in stochastic constitutive models is referred to
\cite{Junker2017} and references cited therein. Solution strategies for problems with
uncertainties in material properties and applied loads are discussed in \cite{Rosic}.

\subsection{Monte Carlo computations using noisy data}

Some preliminary results regarding
the identification of the material paramters for the steel 42CrMo4 were presented in \cite{ShutovKuprin}.
Certain parameters which appear in the material model
can be identified by general considerations. In particular, the elastic constants $k$
and $\mu$ are extracted from the experimental data in the elastic domain. The viscosity
parameters $\eta$ and $m$ can be obtained from a series of uniaxial tension tests with
different loading rates (cf. \cite{ShutovKuprin}).
The pre-identified material parameters are summarized in Tab. \ref{preident}.
The remaining material parameters describe the nonlinear isotropic and
kinematic hardening. They are packed now into the vector
$\vec{p} = (\gamma, \beta, c_1, c_2, \varkappa_1, \varkappa_2)^{\text{T}}$.
Since the mechanisms of isotropic and kinematic hardening are active
at the same time, the corresponding parameters must be
identified simultaneously (cf. \cite{Broggiato}, \cite{ShutovKreissig2010})

\begin{table}[h]
\caption{Pre-identified parameters for the steel 42CrMo4}
\begin{tabular}{| l l l l l |}
\hline
      $k$ [MPa] &  $\mu$ [MPa] & $\eta$ [s]      & $m$ [-] & $K$ [MPa]  \\ \hline
    135 600     & 52000        &  $5 \cdot 10^5$ & 2.26    &  335       \\ \hline
\end{tabular} \\
\label{preident}
\end{table}

In this subsection we demonstrate a procedure for numerical estimation
of the sensitivity of $\vec{p}$ with respect to the measurement errors.
The unknown material parameters are identified using the optimization problem \eqref{OptimizProblem2}.
In this problem, the real experimental data $Exp_i$ are replaced by the noisy data $Exp_i + Noise_i$.
The stochastic model of noise is given by Eq.  \eqref{TwoSourseModel}.
This noise corresponds to two sources of experimental errors: correlated and non-correlated.
Since the mathematical expectation of $Noise_i$ is zero,
the covariance matrix of the noise is given by

\begin{equation}\label{CovarianceMatrix}
\mathbf{Cov}_{i j} = \text{cov}(Noise_i, Noise_j) = \text{E} (Noise_i \cdot Noise_j)=
\sigma_1^2 \delta_{i j} + \sigma_2^2 Exp_i \cdot Exp_j / (\max\limits_{k} | Exp_k |)^2.
\end{equation}
It follows from \eqref{CovarianceMatrix} that the noisy data $Exp_i + Noise_i$ are correlated and exhibit different variations.
The most common optimization procedure, based on the minimization of the error functional with $\mathbf{W} = \text{diag}(1,1,...,1)$,
does not provide most stable results.
In fact, the least square optimization is expected to yield more stable results
when the target values $\mathbf{W}^{1/2} \ (\overrightarrow{\text{Exp}} + \overrightarrow{\text{Noise}})$ are
not correlated and exhibit the same variance (cf. Section 4.6 in reference \cite{Beck}).
Thus, a reasonable choice of the weighting matrix would be
\begin{equation}\label{BestChoice}
\mathbf{W} = \mathbf{Cov}^{-1}.
\end{equation}
Along with \eqref{BestChoice}, we also consider to alternatives: $\mathbf{W} = \text{diag} (1,1,...,1)$ and
$\mathbf{W}_{ij} =  \delta_{ij} / \mathbf{Cov}_{i j}$. Here, $\delta_{ij}$ is the Kronecker delta.
The last choice of $\mathbf{W}$ is a certain approximation of $\mathbf{Cov}^{-1}$ which is still a diagonal matrix.
It exactly coincides with $\mathbf{Cov}^{-1}$ if the noise is not correlated (i.e. if the matrix $\mathbf{Cov}$
 is diagonal).

\begin{figure}\centering
\psfrag{A}[m][][1][0]{$\mathbf{W}  = \text{diag}(1,1,...,1)$}
\psfrag{B}[m][][1][0]{$\mathbf{W}_{ij} =  \delta_{ij} / \mathbf{Cov}_{i j}$}
\psfrag{C}[m][][1][0]{$\mathbf{W}  = \mathbf{Cov}^{-1}$}
\scalebox{0.90}{\includegraphics{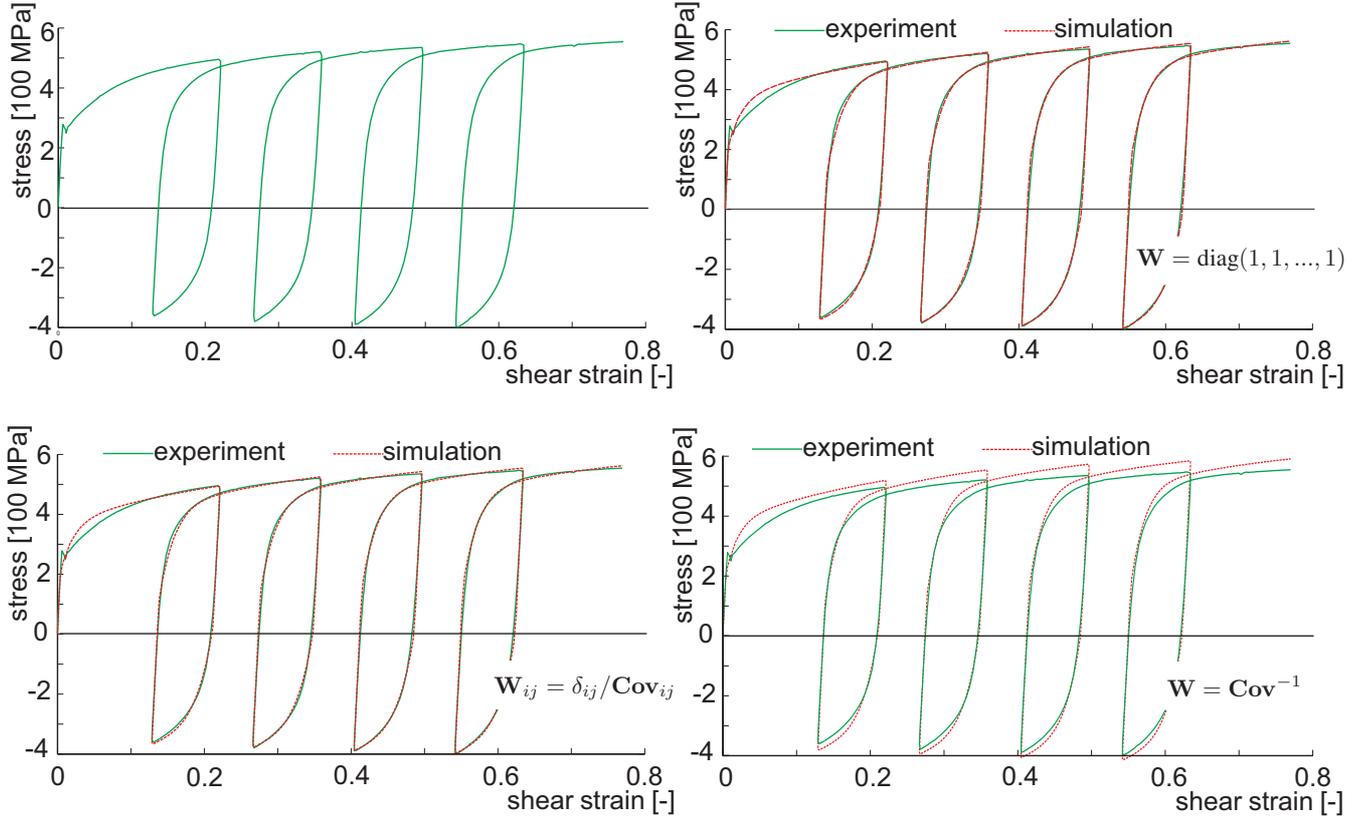}}
\caption{Experiment on non-monotonic torsion of thin-walled specimens made from steel 42CrMo4.
Top left: experimental results from \cite{ShutovKuprin}. Top right:
results of optimization with $\mathbf{W}  = \text{diag}(1,1,...,1)$.
Bottom left: results of optimization with $\mathbf{W}_{ij} =  \delta_{ij} / \mathbf{Cov}_{i j}$.
Bottom right:  results of optimization with $\mathbf{W}  = \mathbf{Cov}^{-1}$. \label{fig2}}
\end{figure}

By $\vec{p}_{\ast}$ we denote the identified parameters for the experimental data
without additional noise. They are summarized in Tab. \ref{table_noisefree}.
The corresponding simulation results are compared with the experimental data in Fig. \ref{fig2}.
As can be seen from the figure, a good correspondence between the simulation and experiment
is observed for the strategies with $\mathbf{W} = \text{diag} (1,1,...,1)$ and
$\mathbf{W}_{ij} =  \delta_{ij} / \mathbf{Cov}_{i j}$. As can be seen from Tab. \ref{table_noisefree},
both strategies provide similar results.  On the other hand, a deviation from the experiment is somewhat larger for the strategy
with $\mathbf{W} = \mathbf{Cov}^{-1}$.

For the Monte Carlo simulations a total number of $N_{noise} = 10 000$ instances of noisy data were considered.
Since the regular error
estimation of the Monte Carlo method is $C/ \sqrt{N_{noise}}$ with a certain constant $C$, the relative error
in the computed size of the parameter cloud is expected to be less than $1 \%$.

The stochastic parameter of the noise \eqref{TwoSourseModel} are $\sigma_1 = 10$ MPa, $\sigma_2 = 5$ MPa.
In order to speed up the Monte Carlo sampling, the model response is
linearized near $\vec{p}_{\ast}$ according to \eqref{Linearization}.
For $j-th$ instance of noise, the corresponding parameter set $\vec{p}^{\ (j)}$ is identified using the
analytical solution \eqref{AnalytSolution}.
In order to give an impression about the distribution of the parameters, the variance
of the normalized parameters is provided in Tab. \ref{table_variance}.
The results indicate that the parameters $\varkappa_1$ and $\varkappa_2$
are much more insensitive to the experimental noise than the parameters $\gamma$ and $\beta$.

\begin{table}[h]
\caption{Identified material parameters for noise-free experimental data for the steel 42CrMo4}
\begin{tabular}{| l | l l l l l l |}
\hline
    & $\gamma_{\ast}$ [MPa] &  $\beta_{\ast}$ [-] & ${c_1}_{\ast}$ [MPa] & ${c_2}_{\ast}$ [MPa] & ${\varkappa_1}_{\ast}$ [1/MPa] & ${\varkappa_2}_{\ast}$ [1/MPa]  \\ \hline
$\mathbf{W}  = \mathbf{Cov}^{-1}$                     & 435.22  &   2.625 & 1 661.7  &   24 672 & 0.003810  &   0.004282  \\ \hline
$\mathbf{W}  = \text{diag}(1,1,...,1)$                & 321.92  &    2.003 & 1 488.4  &   20 512 & 0.004087  &   0.004526  \\ \hline
$\mathbf{W}_{ij} =  \delta_{ij} / \mathbf{Cov}_{i j}$ & 312.60  &    1.913 & 1 505.5  &   20 687 & 0.004089  &   0.004516   \\ \hline
\end{tabular} \\
\label{table_noisefree}
\end{table}

\begin{table}[h]
\caption{Variances of normalized material parameters pertaining to stochastic model \eqref{TwoSourseModel} of noise}
\begin{tabular}{| l | l l l l l l |}
\hline
    & Var$\big(\frac{\displaystyle \gamma}{\displaystyle  \gamma_{\ast}}\big)$  &  Var$\big(\frac{\displaystyle  \beta}{\displaystyle  \beta_{\ast}} \big)$
     & Var$\big(\frac{\displaystyle  c_1}{\displaystyle  {c_1}_{\ast}}  \big)$
     & Var$\big(\frac{\displaystyle  c_2}{\displaystyle  {c_2}_{\ast}} \big)$  & Var$\big(\frac{\displaystyle  \varkappa_1}{\displaystyle  {\varkappa_1}_{\ast}}  \big)$
      & Var$\big(\frac{\displaystyle  \varkappa_2}{\displaystyle  {\varkappa_2}_{\ast}}  \big)$   \\ \hline
$\mathbf{W}  = \mathbf{Cov}^{-1}$                     & 0.00434  &   0.00714 & 0.00132  &   0.000796 & 0.000109  &   0.000187  \\ \hline
$\mathbf{W}  = \text{diag}(1,1,...,1)$                & 0.00740  &    0.0154 & 0.00194  &   0.00152 & 0.000234  &   0.000285  \\ \hline
$\mathbf{W}_{ij} =  \delta_{ij} / \mathbf{Cov}_{i j}$ & 0.00772  &    0.0171 & 0.00193  &   0.00153 & 0.000242  &   0.000283   \\ \hline
\end{tabular} \\
\label{table_variance}
\end{table}

The size of the parameter cloud is then defined as the average distance between $\vec{p}_{\ast}$ and $\vec{p}^{\ (j)}$
\begin{equation}\label{SizeParamCloud}
Size = \frac{1}{N_{noise}} \sum_{j=1}^{N_{noise}} \text{dist}^{\mathbf{F}}(\vec{p}_{\ast}, \vec{p}^{\ (j)}).
\end{equation}

In order to define the distance between two sets of parameters, a
suitable deformation history is needed. In this study we consider two different histories in the time interval
$t \in [0,4]$ ($t$ is a non-dimensional loading parameter here).
In a general case we have
\begin{equation}\label{GeneralHistory}
\mathbf F (t) = \overline{\mathbf F^{\prime} (t)},
\end{equation}
where $\mathbf F^{\prime} (t)$ is a linear interpolation between
certain key-points $\mathbf F_1$, $\mathbf F_2$, $\mathbf F_3$, and $\mathbf F_4$:
\begin{equation*}\label{loaprog}
\mathbf F^{\prime} (t) :=
\begin{cases}
    (1 - t) \mathbf F_1  + (t) \mathbf F_2 \quad \text{if} \ t \in [0,1] \\
    (2 - t) \mathbf F_2  + (t-1) \mathbf F_3 \quad \text{if} \ t \in (1,2] \\
     (3 - t) \mathbf F_3  + (t-2) \mathbf F_4 \quad \text{if} \ t \in (2,3] \\
     (4 - t) \mathbf F_4  + (t-3) \mathbf F_1 \quad \text{if} \ t \in (3,4]
\end{cases}.
\end{equation*}
For the 1st history we employ the following key-points
\begin{equation*}
\mathbf F_1 :=\mathbf 1, \
\mathbf F_2 :=1.2 \ \mathbf{e}_1 \otimes \mathbf{e}_1 + (1.2)^{-1/2} \ (\mathbf{e}_2 \otimes \mathbf{e}_2 + \mathbf{e}_3 \otimes \mathbf{e}_3),
\end{equation*}
\begin{equation*}
\mathbf F_3 := \mathbf 1, \
\mathbf F_4 := 1.2 \ \mathbf{e}_2 \otimes \mathbf{e}_2 +  (1.2)^{-1/2} \ (\mathbf{e}_1 \otimes \mathbf{e}_1 + \mathbf{e}_3 \otimes \mathbf{e}_3),
\end{equation*}
and for the 2nd history we have
\begin{equation*}
\mathbf F_1 :=\mathbf 1, \
\mathbf F_2 := 1.2 \ \mathbf{e}_1 \otimes \mathbf{e}_1 + (1.2)^{-1/2} \ (\mathbf{e}_2 \otimes \mathbf{e}_2 + \mathbf{e}_3 \otimes \mathbf{e}_3),
\end{equation*}
\begin{equation*}
\mathbf F_3 := \mathbf 1 + 0.2 \mathbf{e}_1 \otimes \mathbf{e}_2, \
\mathbf F_4 := 1.2 \ \mathbf{e}_2 \otimes \mathbf{e}_2 + (1.2)^{-1/2} \ (\mathbf{e}_1 \otimes \mathbf{e}_1 + \mathbf{e}_3 \otimes \mathbf{e}_3).
\end{equation*}

The sizes of the parameter clouds are summarized in Table \ref{tab1}.
The simulation results indicate that the strategy with $\mathbf{W}  = \mathbf{Cov}^{-1}$ yields
parameters, which are most stable with respect to the considered noise.
On the other hand, strategies with $\mathbf{W}  = \text{diag}(1,1,...,1)$ and
$\mathbf{W}_{ij} =  \delta_{ij} / \mathbf{Cov}_{i j}$ are almost equivalent to each other
regarding stability of the parameter identification.
Another important conclusion is that the specific choice of the deformation history
is not very important in defining the distance function \eqref{PhysBasedDist}.

\begin{table}[h]
\caption{Sizes (in MPa) of the parameter cloud in terms of the mechanics-based distance}
\begin{tabular}{| l | l l |}
\hline
    & 1st deformation history &  2nd deformation history  \\ \hline
$\mathbf{W}  = \mathbf{Cov}^{-1}$ & 5.522  &   5.562   \\ \hline
$\mathbf{W}  = \text{diag}(1,1,...,1)$ & 7.739  &  7.745   \\ \hline
$\mathbf{W}_{ij} =  \delta_{ij} / \mathbf{Cov}_{i j}$ & 7.687  &    7.682   \\ \hline
\end{tabular} \\
\label{tab1}
\end{table}

\section{Discussion and conclusion}

A simple mechanics-based definition of metric in the space of material parameters is introduced (see Eq. \eqref{PhysBasedDist}).
In contrast to the formal use of the Euclidean norm, this metric
accounts for the importance of each material parameter for the stress response.
The metric is invariant under re-parametrization of the material model.
An interesting conclusion is that the specific choice of the deformation history
has only a minor impact on the results of computations (see Tab. \ref{tab1}).

A strain-controlled loading is implemented in \eqref{PhysBasedDist}
to define a distance between two sets of material parameters. This choise
is reasonable for models of finite strain plasticity and viscoplasticity.
Dealing with models of creep \cite{AltenbachNaumenko}
(including creep damage), a stress-controlled loading can be used instead.

A \emph{local} strain history is considered to define the metric (see Eq. \eqref{PhysBasedDist}).
This definition can be naturally generalized by considering a representative
boundary value problem. The most reasonable results are expected when this
boundary value problem would be close to a specific application.

For each instance of the noisy data, the corresponding parameters
are identified using the linearized problem, where a closed-form solution is available.
Therefore, the presented  approach is computationally efficient.
In a more general case of a large noise, the  assumption \eqref{Linearization} must be dropped
and a straightforward solution of the optimization problem will be needed.

In case of a correlated noise (cf. stochastic model \eqref{TwoSourseModel}), a good stability of the identified parameters
can be achieved by using non-diagonal weighting matrix $\mathbf{W}$.
Unfortunately, there is a certain conflict between the accuracy in the description
of the experimental data and the stability of the identified parameters with respect
to the experimental errors. Thus, the strategy with $\mathbf{W}  = \mathbf{Cov}^{-1}$, which
provides most stable results, yields larger deviation of the computed stress response from the experimental data.
Probably, while solving practical problems, a compromise between the stability and accuracy of the parameter
identification needs to be found. This compromise should be based on the a-priory knowledge of the stochastic
parameters of the experimental error.

The presented method of estimating the sensitivity of the material parameters with respect to the experimental errors
can be useful in various situations. Basically, it can be employed to assess the quality of a certain identification
procedure. When dealing with experimental data pertaining to different types of experiments, like tension-compression or non-monotonic torsion, a problem arises
of how to combined these data in a single error functional. If a realistic model of stochastic noise for different experiments is available,
the method can be used to define suitable weighting coefficients.
Application of the mechanics-based metric to optimal experimental
design problems (cf. \cite{Herzog}) is
also promising.

Acknowledgement. The financial support provided
by the RFBR (grant number 17-08-01020) is acknowledged.

\end{document}